\documentclass[aps,prl,showpacs,amsmath,amssymb,superscriptaddress,reprint,10pt]{revtex4-1}
\usepackage{bm}
\usepackage{graphicx}
\usepackage{xcolor}
\usepackage{url}
\usepackage[normalem]{ulem}
\usepackage{times}
\usepackage{bbm}
\usepackage{mathrsfs} 
\usepackage[colorlinks=true,breaklinks=true,allcolors=blue]{hyperref}

\definecolor{romain}{rgb}{1,.4,.3}

\definecolor{david}{rgb}{0,0,1}

\definecolor{michael}{rgb}{0,.8,.5}

\newcommand\Sphere{{\mathbbm{S}}}
\newcommand\XX{{\mathbbm{X}}}
\newcommand\YY{{\mathbbm{Y}}}

\newcommand\ZZ{{\mathbbm{Z}}}
\newcommand\RR{{\mathbbm{R}}}

\newcommand\id{{\mathbbm{1}}}

\newcommand\dd{{\mathrm{d}}}
\newcommand\ee{{\mathrm{e}}}

\newcommand{\vertl}{\left\lvert}
\newcommand{\vertr}{\right\rvert}

\newcommand{\intt}{\int_{0}^{t}}

\newcommand{\derp}[2]{\dfrac{\partial #1}{\partial #2}}
\newcommand{\ds}{\mathrm{d}s}
\newcommand{\dt}{\mathrm{d}t}
\newcommand{\psii}{\psi^{i}}
\newcommand{\deltai}{\delta^{i}}

\newcommand{\J}{J_\Lambda}
\newcommand{\q}{q}
\newcommand{\p}{p}

\begin{document}

\title{Spreading of Perturbations in Long-Range Interacting Classical Lattice Models}  

\author{David M\'etivier} 
\affiliation{Universit\'e de Lyon, {\'E}cole Normale Sup\'erieure de Lyon, 46 All\'ee d'Italie, 69364 Lyon cedex 07, France} 
\affiliation{National Institute for Theoretical Physics (NITheP), Stellenbosch 7600, South Africa} 
\affiliation{Institute of Theoretical Physics,  University of Stellenbosch, Stellenbosch 7600, South Africa}

\author{Romain Bachelard} 
\affiliation{Instituto de F\'{\i}sica de S\~ao Carlos, Universidade de S\~ao Paulo, 13560-970 S\~ao Carlos, SP, Brazil}

\author{Michael Kastner} 
\email{kastner@sun.ac.za} 
\affiliation{National Institute for Theoretical Physics (NITheP), Stellenbosch 7600, South Africa} 
\affiliation{Institute of Theoretical Physics,  University of Stellenbosch, Stellenbosch 7600, South Africa}

\date{\today}

\begin{abstract}
Lieb-Robinson-type bounds are reported for a large class of classical Hamiltonian lattice models. By a suitable rescaling of energy or time, such bounds can be constructed for interactions of arbitrarily long range. The bound quantifies the dependence of the system's dynamics on a perturbation of the initial state. The effect of the perturbation is found to be effectively restricted to the interior of a causal region of logarithmic shape, with only small, algebraically decaying effects in the exterior. A refined bound, sharper than conventional Lieb-Robinson bounds, is required to correctly capture the shape of the causal region, as confirmed by numerical results for classical long-range $XY$ chains. We discuss the relevance of our findings for the relaxation to equilibrium of long-range interacting lattice models.
\end{abstract}


\maketitle 

In many nonrelativistic lattice systems, and despite the absence of Lorentz covariance, physical effects are mostly restricted to a causal region, often in the shape of an effective ``light cone,'' with only tiny effects leaking out to the exterior. The technical tool, known as Lieb-Robinson bounds \cite{LiebRobinson72,NachtergaeleSims10,*Review}, to quantify this statement in a quantum mechanical context is an upper bound on the norm of the commutator $\left[O_A(t),O_B(0)\right]$, where $O_A(0)$ and $O_B(0)$ are operators supported on the subspaces of the Hilbert space corresponding to nonoverlapping regions $A$ and $B$ of the lattice. The importance of such a bound lies in the fact that a multitude of physically relevant results can be derived from it. Examples are bounds on the creation of equal-time correlations \cite{NachtergaeleOgataSims06}, on the transmission of information \cite{BravyiHastingsVerstraete06}, and on the growth of entanglement \cite{EisertOsborne06}, 
the exponential spatial decay of correlations in the ground state of a gapped system \cite{HastingsKoma06}, or a Lieb-Schultz-Mattis theorem in higher dimensions \cite{Hastings04}. Experimental observations related to Lieb-Robinson bounds have also been reported \cite{Cheneau_etal12,*Langen_etal13}.

The original proof by Lieb and Robinson \cite{LiebRobinson72} requires interactions of finite range. An extension to power-law-decaying long-range interactions has been reported in Refs.\ \cite{HastingsKoma06,NachtergaeleOgataSims06}. In this case the effective causal region is no longer cone shaped, and the spatial propagation of physical effects is not limited by a finite group velocity \cite{EisertvdWormManmanaKastner13}. For ``strong long-range interactions,'' i.e., when the interaction potential decays proportionally to $1/r^\alpha$ with an exponent $\alpha$ smaller than the lattice dimension $d$, the theorems in \cite{HastingsKoma06,NachtergaeleOgataSims06} do not apply and no Lieb-Robinson-type results are known. This fact nicely fits into the larger picture that, for $\alpha\leq d$, the behavior of long-range interacting systems often differs substantially from that of short-range interacting systems. Examples of such differences include nonequivalent equilibrium statistical ensembles and negative response functions \cite{LynWood68,*Thirring70,*ElHaTur00,*TouElTur04,*Kastner10,*KastnerJSTAT10}, or the occurrence of quasistationary states whose lifetimes diverge with the system size \cite{AnRu95,*Kastner11,*Kastner12,CamDauxRuf09}. The latter is a dynamical phenomenon, and it has been conjectured in \cite{BachelardKastner13} that some of its properties are universal and in some way connected to Lieb-Robinson bounds. 

In most cases the peculiarities of long-range interacting systems have been investigated in the framework of classical Hamiltonian systems \cite{CamDauxRuf09}, but little is known about Lieb-Robinson bounds in classical mechanics. Exceptions are restricted to specific models with nearest-neighbor interactions \cite{Marchioro_etal78,Butta_etal07,*RazSims09}. In the context of classical Hamiltonian mechanics, a Lieb-Robinson bound is an upper bound on the norm of the Poisson bracket $\left\{f_A(0),g_B(t)\right\}$, where $f_A(0)$ and $g_B(0)$ are phase space functions supported only on the subspaces of phase space corresponding to the nonoverlapping regions $A$ and $B$ of the lattice, respectively. The physical meaning of the norm of this Poisson bracket becomes evident from an expression put forward in \cite{Marchioro_etal78},
\begin{equation}\label{e:Poisson}
\left\lvert\left\{f_A(0),g_B(t)\right\}\right\rvert \leq |A||B|\Vert\nabla f\rVert_\infty \Vert\nabla g\rVert_\infty u_{AB}(t)
\end{equation}
where $\Vert\nabla f\rVert_\infty$ and $\Vert\nabla g\rVert_\infty$ are the (bounded) maxima of all the partial derivatives of $f$ and $g$ with respect to the phase space coordinates, and
\begin{equation}\label{e:uij}
u_{AB}(t) = 4\max_{\substack{i\in A\\j\in B}}\left\{\left\lvert\frac{\partial p_j(t)}{\partial p_i(0)}\right\rvert\!,\left\lvert\frac{\partial q_j(t)}{\partial p_i(0)}\right\rvert\!,\left\lvert\frac{\partial p_j(t)}{\partial q_i(0)}\right\rvert\!,\left\lvert\frac{\partial q_j(t)}{\partial q_i(0)}\right\rvert\right\}\!.
\end{equation}
The partial derivatives on the right-hand side of \eqref{e:uij} quantify the effect that a variation of the initial momentum or position $p_i(0)$, $q_i(0)$ at the lattice site $i$ have on the time-evolved momentum or position $p_j(t)$, $q_j(t)$ at the lattice site $j$. A classical Lieb-Robinson bound is therefore a measure for the spreading in time and space of an initial perturbation, with potential applications to a broad range of physical processes, including heat conduction, signal transmission, transfer of energy, or the approach to equilibrium.

In this Letter we study the spreading in time and space of initial perturbations in classical long-range interacting lattice models. We have proved a Lieb-Robinson-type result, providing an upper bound on $u_{AB}(t)$ in \eqref{e:uij} [and hence on the Poisson bracket \eqref{e:Poisson}] for a broad class of classical long-range interacting lattice models in arbitrary spatial dimension. Dipolar interactions in condensed matter systems are the prime example of such long-range interacting lattice systems \cite{Miloshevich_etal13}, but many other examples exist \cite{Bachelard_etal10}. To avoid the rather technical notation of the general result \footnote{See Supplemental Material.}, we present the main result in this Letter for a specific class of systems, namely classical $XY$ models in $d$ spatial dimensions with pair interactions that decay like a power law $1/|i-j|^\alpha$ with the (1-norm) distance $|i-j|$ between lattice sites $i$ and $j$. The value of the exponent $\alpha$ determines the range of the interaction, from mean-field-type (distance-independent) interactions at $\alpha=0$ to nearest-neighbor couplings in the limit $\alpha\to\infty$.

Our study focuses on the influence of the interaction range on the spreading of perturbations, and we find pronounced quantitative and qualitative changes upon variation of $\alpha$. Different from systems with finite-range interactions, the effect of an initial perturbation is found to be effectively restricted to the interior of a causal region of logarithmic shape, with algebraically small effects in the exterior. Similar to the short-range case, such a bound can be used to rigorously control finite-size effects in simulations of lattice models, exclude information transmission above a certain measurement resolution in the exterior of the effective causal region, and much more. Our analytical results are supplemented by numerical simulations of the time evolution of a long-range interacting $XY$ chain. Besides confirming the validity of the bound, the numerical results reveal that the refined version of our bound, sharper than conventional Lieb-Robinson-type bounds, is required in order to correctly capture the shape of the propagation front.

{\em $\alpha XY$ model.---}This model consists of classical $XY$ spins (or rotors) attached to the sites $i\in\Lambda$ of a $d$-dimensional hypercubic lattice $\Lambda\subset\ZZ^d$. The phase space of a single rotor is $X_i=\Sphere^1\times\RR$, allowing to parametrize each rotor by an angular variable $q_i\in\Sphere^1$ and by its angular momentum $p_i\in\RR$. On the phase space $\XX=X_1\times\dotsb\times X_{|\Lambda|}$ of the total system we define the Hamiltonian function
\begin{equation}\label{e:H}
H=\sum_{i\in\Lambda}\frac{p_i^2}{2}-\frac{J_\Lambda}{2}\sum_{\substack{i,j\in\Lambda\\i\neq j}} \frac{\cos(q_i-q_j)}{|i-j|^\alpha}.
\end{equation}
For $\alpha\leq d$, the second sum on the right-hand side of \eqref{e:H} is superextensive, i.e., asymptotically for large lattices it grows faster than linearly with the number $|\Lambda|$ of lattice sites. Our proof of a Lieb-Robinson bound requires the Hamiltonian to be extensive. We enforce extensivity also for $\alpha\leq d$ by allowing the coupling constant to depend explicitly on the lattice,
\begin{equation}\label{e:JLambda}
J_\Lambda=J \Big/\sup_{i\in\Lambda}\sum_{j\in\Lambda\setminus\{i\}}\frac{1}{|i-j|^{\alpha}},
\end{equation}
where $J$ is a real constant \footnote{Introducing such a normalization is actually a harmless procedure and amounts to a rescaling of time. For a ``physical'' Hamiltonian without such a rescaling, the bounds we derived hold therefore in rescaled time $t J_\Lambda$. For finite systems, no rescaling is necessary.}.

{\em Classical long-range Lieb-Robinson bound.---}Upper bounds on the partial derivatives on the right-hand side of \eqref{e:uij} are given by
\begin{subequations}
\begin{align}\displaybreak[2]
\left\lvert\frac{\partial q_j(t)}{\partial q_i(0)}\right\rvert &\leq \frac{\sum\limits_{n=1}^\infty
\dfrac{U_{n}^{ij} t^{2n}}{(2n)!}}{|i-j|^\alpha}\leq \frac{\cosh(v t)-1}{|i-j|^\alpha}=:B_{ij}^{qq}(t),\label{e:pq1}\\
\left\lvert\frac{\partial q_j(t)}{\partial p_i(0)}\right\rvert&\leq\frac{\sum\limits_{n=1}^\infty
\dfrac{U_{n}^{ij} |t|^{2n+1}}{(2n+1)!}}{|i-j|^\alpha}\leq\frac{\sinh|vt|-|vt|}{v |i-j|^\alpha}=:B_{ij}^{qp}(t),\label{e:pq2}\\
\left\lvert\frac{\partial p_j(t)}{\partial q_i(0)}\right\rvert&\leq\frac{\sum\limits_{n=1}^\infty
\dfrac{U_{n}^{ij} |t|^{2n-1}}{(2n-1)!}}{|i-j|^\alpha}\leq\frac{v \sinh|vt|}{|i-j|^\alpha}=:B_{ij}^{pq}(t),\label{e:pq3}\\
\left\lvert\frac{\partial p_j(t)}{\partial p_i(0)}\right\rvert &\leq \frac{\sum\limits_{n=1}^\infty
\dfrac{U_{n}^{ij} t^{2n}}{(2n)!}}{|i-j|^\alpha}\leq \frac{\cosh(v t)-1}{|i-j|^\alpha}=:B_{ij}^{pp}(t).\label{e:pq4}
\end{align}
\end{subequations}
The positive coefficients $U_{n}^{ij}$ are defined recursively by
\begin{equation}\label{e:recursion}
U_{n+1}^{ij}=
     \begin{cases}
       \lvert J\rvert U_{n}^{ij}+\lvert\J\rvert U_{n}^{ii}+C_{ij}U_{n}^\text{max} & \text{for $i\neq j$,} \\
        \lvert J\rvert U_{n}^{ii}+C_{ii}U_{n}^\text{max} & \text{for $i=j$,}
     \end{cases}
\end{equation}
with $U_1^{ij}=\lvert\J\rvert$ and $U_1^{ii}=\lvert J\rvert$, and we use the constants
\begin{gather}
U_{n}^\text{max}=\sup_{i,j\in\Lambda}U_{n}^{ij},\qquad v=\sup_{i,j\in\Lambda}\sqrt{\lvert J+J_\Lambda\rvert+C_{ij}},\label{e:v}\\
C_{ij} =\begin{cases}
       \displaystyle \lvert J_\Lambda\rvert\sum_{k\in\Lambda\setminus\{i,j\}}\frac{|i-j|^\alpha}{|i-k|^\alpha|j-k|^\alpha} & \text{for $i\neq j$,} \\
        \displaystyle \lvert J_\Lambda\rvert\sum_{k\in\Lambda\setminus\{i\}}\frac{1}{|i-k|^{2\alpha}} & \text{for $i=j$.}
     \end{cases}\label{e:Cij}
\end{gather}

The proof of the bounds combines techniques for classical lattices with nearest-neighbor interactions \cite{Marchioro_etal78} with those used for proving Lieb-Robinson bounds for long-range quantum systems \cite{HastingsKoma06,NachtergaeleOgataSims06}, with the additional refinement of allowing lattice-dependent coupling constants. In the thermodynamic limit of infinite lattice size, the quantity $v$ in \eqref{e:v} remains finite and hence the bounds remain meaningful \cite{Note1}.

All the bounds on the right-hand side of \eqref{e:pq1}--\eqref{e:pq4}, and therefore also the norm of the Poisson bracket \eqref{e:Poisson}, grow exponentially for large times $|t|$, and decay as a power law with the distance $|i-j|$. For some $\epsilon>0$, the effect of a perturbation is therefore smaller than $\epsilon$ outside a region in the $(|i-j|,t)$ plane specified, for large $|t|$, by
\begin{equation}\label{e:causalregion}
v|t|>\ln(2\epsilon)+\alpha\ln|i-j|.
\end{equation}
This effective causal region, in which the effect of an initial perturbation is non-negligible, has a logarithmic shape, and differs in this respect from the linear (cone-shaped) region derived in \cite{Marchioro_etal78} for short-range interactions. As a consequence, no finite group velocity limits the spreading of perturbations in long-range interacting lattices, and supersonic propagation can occur. For a refined understanding of the spatiotemporal behavior, we go beyond the usual Lieb-Robinson-type estimates and consider the sharper bounds in \eqref{e:pq1}--\eqref{e:pq4}, where the coefficients $U_n^{ij}$ introduce an additional spatial dependence. The functional form of these bounds will be illustrated below, and we also show that only the sharper bounds correctly capture the shape of the propagation front.

The bounds \eqref{e:pq1}--\eqref{e:pq4} remain valid, with only minor modifications of the parameters and prefactors, under broad generalizations, including arbitrary graphs $\Lambda$, multidimensional single-particle phase spaces $X_i$, many-particle interactions, and very general forms of the interaction potential \cite{Note1}. In the remainder of this Letter we subject the bounds \eqref{e:pq1}--\eqref{e:pq4} to a reality check, in the sense of testing their tightness and whether the form of the propagation front obtained numerically is faithfully reproduced. 

\begin{figure}\centering
\includegraphics[width=\linewidth]{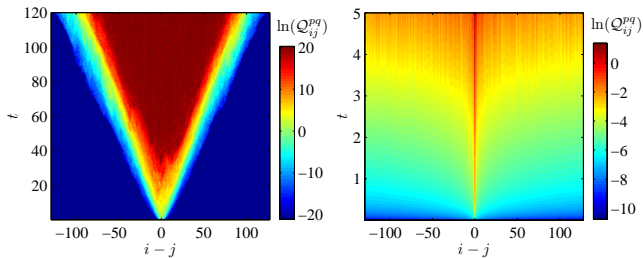}
\caption{\label{f:horizon}%
Illustration of the spatiotemporal behavior of perturbations. The contour plots show $\ln\mathcal{Q}_{ij}^{pq}$ as a function of the distance $|i-j|$ and time $t$, for chains of length $N=256$. Left: for the $XY$ chain with nearest-neighbor interactions, the effect of a perturbation is restricted to the interior of a cone-shaped region. Right: for the $\alpha XY$ chain with $\alpha=1/2$, the contours spread faster than linearly in space, as expected from \eqref{e:causalregion}, illustrating supersonic propagation.
}%
\end{figure}

{\em Numerics.---}We consider the $\alpha XY$ model \eqref{e:H} on a ring, i.e., a one-dimensional chain of $N=|\Lambda|$ sites with periodic boundary conditions. The partial derivatives in \eqref{e:pq1}--\eqref{e:pq4} are approximated by difference quotients,
\begin{equation}\label{e:diff_quotient}
\frac{\partial p_j(t)}{\partial q_i(0)}\approx\frac{\tilde{p}_j(t)-p_j(t)}{\delta q_i}=:Q_{ij}^{pq}(t),
\end{equation}
and similarly for the other derivatives. Here, $p_j(t)=p_j(t,p_1(0),\dotsc,p_{N}(0), q_1(0),\dotsc,q_{N}(0))$ is the time-evolved momentum obtained by starting from a certain initial condition, and $\tilde{p}_j(t)=p_j(t,p_1(0),\dotsc,p_{N}(0),q_1(0),\dotsc,q_i(0)+\delta q_i,\dotsc,q_{N}(0))$ is for a similar initial condition, but with the $i$th initial position shifted by some small $\delta q_i$. The time-evolved momenta $p_j$ and $\tilde{p}_j$ are obtained by numerically integrating Hamilton's equations using a sixth-order symplectic integrator \cite{McLachlanAtela92}. The numerical results for $Q_{ij}^{pq}$ fluctuate strongly in time, obscuring the overall trend of the spreading. To reduce the fluctuating background, we compute the difference quotient \eqref{e:diff_quotient} for 20 different (pairs of) initial conditions; details regarding the choice of initial conditions are given in the Supplemental Material \cite{Note1}.  Since our aim is to compare the numerical results to the upper bounds \eqref{e:pq1}--\eqref{e:pq4}, we select, for any fixed time $t$ and lattice sites $i$ and $j$, the largest of the 20 $Q_{ij}^{pq}$-values. The resulting maximum is denoted by $\mathcal{Q}_{ij}^{pq}$, and its time- and distance-dependence is shown in Fig.\ \ref{f:horizon}. The plots illustrate the supersonic propagation of perturbations in the presence of long-range interactions, as expected from the inequality \eqref{e:causalregion}. 

\begin{figure}\centering
\includegraphics[width=\linewidth]{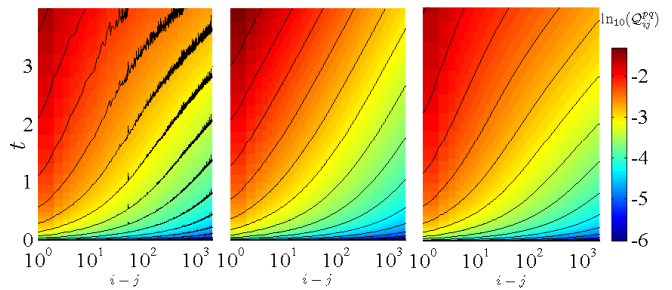}
\caption{\label{f:horizonfit}%
Numerical data and fits of the spreading of perturbations in the $\alpha XY$ chain with $\alpha=1/2$. The contour plots show $\ln\mathcal{Q}_{ij}^{pq}$ as a function of the distance $|i-j|$ on a logarithmic scale and time $t$ on a linear scale. Left: numerical data for a chain of length $N=4096$. Center: fit of the function $cB_{ij}^{pq}(t/z)$ [based on the weaker bound in \eqref{e:pq3}] to the numerical data of $\mathcal{Q}_{ij}^{pq}(t)$, with fit parameters $c=0.0064$ and $z=1.47$, yielding a residual sum of squares of $0.157$. The contours of the bound are approximately linear for large $|i-j|$ and $t$, but this does not correctly capture the actual behavior of the data. Right: As in the center plot, but fitting the parameters $\tilde{c}=21.5$ and $\tilde{z}=11.2$ in the stronger bound \eqref{e:tildefit} and yielding a residual sum of squares as small as $0.0065$.
}%
\end{figure}

For all distances $|i-j|$ and times $t$, the numerical results are smaller than the bounds \eqref{e:pq1}--\eqref{e:pq4} and hence confirm their validity. What is more, the results nicely agree with the functional forms of the bounds, and only the prefactors are overestimated. This observation suggests fitting the function $cB_{ij}^{pq}(t/z)$ to the numerical data of $\mathcal{Q}_{ij}^{pq}(t)$ (and similarly for the other derivatives), with $c$ and $z$ as fit parameters (see left and center plots of Fig.\ \ref{f:horizonfit}). Although the quality of this fit (having a residual sum of squares of $0.157$) is acceptable, it can be improved by about two orders of magnitude by using the fit function
\begin{equation}\label{e:tildefit}
\tilde{B}_{ij}^{pq}(t)=\frac{\tilde{c}}{|i-j|^\alpha}
\sum\limits_{n=1}^\infty
\dfrac{U_{n}^{ij} |t/\tilde{z}|^{2n-1}}{(2n-1)!}
\end{equation}
based on the sharper bound in \eqref{e:pq3}, with fit parameters $\tilde{c}$ and $\tilde{z}$. This fit is of excellent quality, indicating that the distance-dependence of the coefficients $U_n^{ij}$ in \eqref{e:v} appreciably modifies the shape of the propagation front and correctly reproduces the actual spreading.

\begin{figure}\centering
\includegraphics[width=\linewidth]{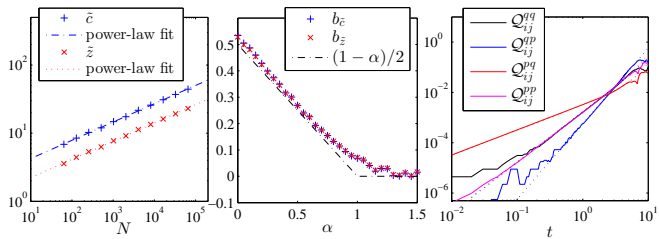}
\caption{\label{f:N-dependence}%
Left and center: system-size dependence of the parameters $\tilde{c}$ and $\tilde{z}$ when fitting \eqref{e:tildefit} to the numerical data of $\alpha XY$ chains, using initial conditions with zero initial momenta. The data for $\alpha=1/2$ in the left plot are well described by the power laws $\tilde{c}\approx2.3 N^{b_{\tilde{c}}}$ and $\tilde{z}\approx1.2 N^{b_{\tilde{z}}}$ with $b_{\tilde{c}}\approx b_{\tilde{z}}\approx0.27$. Right: $b_{\tilde{c}}$ and $b_{\tilde{z}}$ as functions of the exponent $\alpha$. For $\alpha<1$ both are well fitted by the linear function $(1-\alpha)/2$. Right: short-time behavior of the difference quotients $\mathcal{Q}_{ij}^{qq}$, $\mathcal{Q}_{ij}^{qp}$, $\mathcal{Q}_{ij}^{pq}$, and $\mathcal{Q}_{ij}^{qq}$, plotted on a log-log scale. Data are for $\alpha=1/2$ and chain length $N=4096$. The solid data curves display a linear, quadratic, or cubic initial growth, in agreement with the corresponding bounds. Dotted lines are fits of $cB_{ij}(t/z)$, with $c$ and $z$ as fitting parameters.
}%
\end{figure}

The sharper bounds in \eqref{e:pq1}--\eqref{e:pq4} inherit a system-size dependence through the lattice dependence of $J_\Lambda$ and $C_{ij}$ in the coefficients $U_n^{ij}$. As a result, $U_n^{ij}$ (at fixed $n$ and fixed distance $|i-j|$) scales differently with $N$ for the cases $0\leq\alpha<d/2$, $d/2<\alpha<d$, and $\alpha>d$, respectively \cite{Note1}. The switching from one regime to another at $\alpha=d/2$ and $\alpha=d$ nicely coincides with the different scaling regimes of equilibration times observed in \cite{BachelardKastner13}. Additionally to the $N$-dependence inherent to the bound, we find that, for $\alpha\leq d$, the optimal values for the fit parameters $\tilde{c}$ and $\tilde{z}$ in \eqref{e:tildefit} show a strong $N$-dependence, well-captured by a power law $\propto N^{(1-\alpha)/2}$ for both parameters (Fig.\ \ref{f:N-dependence} left and center). This scaling seems to originate from the $N$-dependence of the prefactor $J_\Lambda$ in \eqref{e:H} and \eqref{e:Cij}, and is seen as an indication that the bounds could be further improved. For $\alpha>d$, in contrast, the $N$-dependence of $\tilde{c}$ and $\tilde{z}$ is negligible.

{\em Comparison to the quantum mechanical bound.---} Lieb-Robinson bounds were previously known for long-range interacting {\em quantum}\/ systems \cite{HastingsKoma06,NachtergaeleOgataSims06}. The functional form of these bounds is $c(\ee^{v|t|}-1)/(1+|i-j|)^\alpha$, with a constant $c$ that depends on the observables considered. Asymptotically for long times $t$ and large distances $|i-j|$, this functional form coincides with that of the weaker form of all four classical bounds \eqref{e:pq1}--\eqref{e:pq4}. For small $t$, however, the bounds differ, not only between the classical and the quantum case, but also between the four derivatives bounded in the classical case. The short-time behavior is linear in $t$ for $B_{ij}^{pq}$ in \eqref{e:pq3}, quadratic in $t$ for $B_{ij}^{qq}$ and $B_{ij}^{pp}$ in \eqref{e:pq1} and \eqref{e:pq4}, and cubic in $t$ for $B_{ij}^{qp}$ in \eqref{e:pq2}. The numerical results in Fig.\ \ref{f:N-dependence} (right) confirm that the real short-time dynamics of the $\alpha XY$ model is correctly captured by these different functional forms of the bounds. The quantum mechanical bound, in contrast, increases linearly for short times $t$, independently of the observables considered, although this may not reflect the actual behavior of expectation values in all cases.

{\em Conclusions.---} We have reported Lieb-Robinson-type inequalities, bounding the speed at which a perturbation can travel across the lattice, for a broad class of long-range interacting classical lattice models (including models on arbitrary graphs $\Lambda$, with multidimensional single-particle phase spaces $X_i$, many-particle interactions, and for rather general forms of the interaction potential). By a suitable rescaling, we extended the bounds to arbitrary non-negative long-range exponents $\alpha$, deep into the regime of strong long-range interactions. The weaker bounds on the right-hand side of \eqref{e:pq1}--\eqref{e:pq4} are direct analogs of the quantum mechanical version of Lieb-Robinson bounds for long-range interacting systems \cite{HastingsKoma06,NachtergaeleOgataSims06}. While our numerical results for $\alpha XY$ chains confirm the validity of these bounds, they reveal that the shape of the propagation front is not correctly captured. Only the stronger versions of the bounds in \eqref{e:pq1}--\eqref{e:pq4}, with an additional distance-dependence introduced through the coefficients $U_n^{ij}$, reproduce the functional form of the propagation front. These findings are in contrast to the short-range case, where already the weaker ``conventional'' form of the Lieb-Robinson bound yields the correct, cone-shaped spatiotemporal behavior in agreement with the numerical results.

Since our results apply to arbitrary classical observables, potential applications cover a broad range of dynamical phenomena in long-range interacting classical lattice models, from heat conduction to information transmission, energy transfer, and the approach to equilibrium. In the latter context, different finite-size scaling properties of equilibration times had been observed in the regimes $0\leq\alpha<d/2$, $d/2<\alpha<d$, and $\alpha>d$, respectively \cite{BachelardKastner13}. These three regimes agree precisely with the different scaling regimes of the coefficients $C_{ij}$ that enter and reflect in the bounds \eqref{e:pq1}--\eqref{e:pq4}, providing a theoretical explanation of the numerical observations.

Sharper Lieb-Robinson bounds, similar in spirit to \eqref{e:pq1}--\eqref{e:pq4}, can also be derived for quantum mechanical lattice models with long-range interactions and will be reported in a forthcoming paper.


R.\,B.\ acknowledges support by the Funda\c{c}\~ao de Amparo \`a Pesquisa do Estado de S\~ao Paulo (FAPESP), and the computational support of the N\'ucleo de Apoio a \'Optica e Fot\^onica (NAPOF-USP). M.\,K.\ acknowledges financial support by the National Research Foundation of South Africa via the Incentive Funding and the Competitive Programme for Rated Researchers.


\bigskip

\newpage

\begin{center}
{\bf Supplemental Material}  
\end{center}
\vspace{-3mm}
\appendix
\numberwithin{equation}{section}
\numberwithin{figure}{section}

\section{A. Proof of Equations (5)--(7)}
\setcounter{section}{1}
\setcounter{equation}{0}
\setcounter{figure}{0}

According to Hamilton's equations, the equations of motion corresponding to the Hamiltonian (3) are
\begin{subequations}
\begin{align}
\dfrac{\text{d}}{\text{d} t}p_{j}(t)&=-\J\sum_{\substack{i\in\Lambda\\i\neq j}} \dfrac{\sin\left[ q_{j}(t)-q_{i}(t)\right]}{\vertl i-j\vertr^\alpha},
\label{e:Homega}\\
\dfrac{\mathrm{d}}{\text{d} t}q_{j}(t)&=p_{j}(t).
\label{e:Hq_}
\end{align}
\end{subequations}
Integrating \eqref{e:Homega} and \eqref{e:Hq_} over time $t$ and taking partial derivatives with respect to $q_i(0)$ or $p_i(0)$, one obtains
\begin{subequations}
\begin{align}
\derp{q_j(t)}{q_i(0)}&=\delta_{ij}+\intt\derp{p_j(s)}{q_i(0)}\ds,\label{e:tt}\\
\derp{q_j(t)}{p_i(0)}&=\intt\derp{p_j(s)}{p_i(0)}\ds,\label{e:to}\\
\derp{p_j(t)}{q_i(0)}&=-\intt\J\label{e:ot}\\
\times&\sum_{\substack{k\in\Lambda \\k\neq j}}\left[\frac{\cos[q_j(s)-q_k(s)]}{|j-k|^\alpha}\left(\derp{q_j(s)}{q_i(0)}-\derp{q_k(s)}{q_i(0)}\right)\right]\ds,\nonumber\\
\derp{p_j(t)}{p_i(0)}&=\delta_{ij}-\intt \J\label{e:oo}\\
\times&\sum_{\substack{k\in\Lambda \\k\neq j}}\left[\frac{\cos[q_j(s)-q_k(s)]}{|j-k|^\alpha}\left(\derp{q_j(s)}{p_i(0)}-\derp{q_k(s)}{p_i(0)}\right)\right]\ds.\nonumber
\end{align}
\end{subequations}
These are the four derivatives occurring in (2), and we want to derive upper bounds on their absolute values. Here we show only the derivation of the bound on the absolute value of \eqref{e:tt}. Bounds on the other derivatives can be obtained by the same strategy. For the sake of a compact notation, we introduce the definitions
\begin{subequations}
\begin{align}
\psii&=\left(\psii_{j}\right)_{j\in\Lambda}\qquad\text{with}\quad\psii_{j}(t)=\derp{q_j(t)}{q_i(0)},\label{e:psi}\\
A_{jj}(t)&=-\J\sum_{\substack{k\in\Lambda\\k\neq j}}\left(\dfrac{\cos[q_j(t)-q_k(t)]}{\vertl j-k\vertr^\alpha}\right),\\
A_{jk}(t)&=\J\dfrac{\cos[q_j(t)-q_k(t)]}{\vertl j-k\vertr^\alpha}\quad\text{for $k\neq j$}.\label{e:Ajk}
\end{align}
\end{subequations}
We denote by $A(t)$ the matrix with elements $A_{jk}(t)$, and we define the vectors $\psii=\bigl({\psii_{1}},\dotsc,{\psii_{|\Lambda|}}\bigr)$ and $\deltai=\bigl({\deltai_{1}}\dotsc{\deltai_{|\Lambda|}}\bigr)$. Inserting \eqref{e:ot} into \eqref{e:tt} and expressing the result in terms of the definitions \eqref{e:psi}--\eqref{e:Ajk} yields
\begin{equation}\label{e:dintegral}
\psii_{j}(t)=\delta_{ij}+\intt\int_0^{t_1} \left[A\psii\right]_j\!(t_2)\,\dt_2\,\dt_1,
\end{equation}
where $\left[A\psii\right]_j$ denotes the $j$th component of the vector $A\psii$. 
Integrating by parts we obtain
\begin{equation}
\psii_{j}(t)=\delta_{ij}+\intt(t-t_1) \left[A\psii\right]_j\!(t_1)\,\dt_1 .
\end{equation}
$M$-fold iteration of this formula gives
\begin{multline}
\psii_j(t)=\delta_{ij}+\sum_{m=1}^M \biggl(\intt\int^{t_1}_0\cdots\int^{t_{m-1}}_0\dt_m\cdots\dt_1\\
\times(t-t_1)\cdots(t_{m-1}-t_m)\left[A(t_1)\cdots A(t_m)\deltai\right]_{j}\biggr)\\
+\intt\int^{t_1}_0 \cdots\int^{t_{M}}_0 
(t-t_1)\cdots(t_{M}-t_{M+1})\\
\times\left[A(t_1)\cdots A(t_{M+1})\psii(t_{M+1})\right]_{j}\dt_{M+1}\cdots\dt_1,
\label{e:series}
\end{multline}
where $\delta^i=\left(\delta_{ij}\right)_{j\in\Lambda}$ is a vector of Kronecker deltas. To prove a bound on this series, and hence also its convergence, in the large-$M$ limit, we proceed by constructing an upper bound on (the absolute value of the elements of) the matrix products in \eqref{e:series},
\begin{equation}
  \Biggl\lvert\sum_{k_1,\dotsc,k_{n-1}\in\Lambda}\!\!\!\!\!\!A_{jk_1}(t_1)\cdots A_{k_{n-1}i}(t_{n})\Biggr\rvert\leq
     \begin{cases}
        \dfrac{U_{n}^{ij}}{\vertl i-j\vertr^\alpha} & \text{\!\!\!for $i\neq j$,} \\
        U_{n}^{ii} & \text{\!\!\!for $i=j$,}
     \end{cases}
     \label{e:rec}
\end{equation}
with $U_n^{ij}$ as defined in (6). For the proof of \eqref{e:rec} we require that
\begin{equation}\label{e:repro}
C_{ij}<\infty\qquad\forall i,j\in\Lambda,~i\neq j
\end{equation}
with $C_{ij}$ as defined in (8). We postpone a detailed discussion of this condition to Sec.\ B.1, where we prove that \eqref{e:repro} is satisfied for power law decaying long-range interactions with exponents $\alpha\geq0$. Provided \eqref{e:repro} holds, we can prove \eqref{e:rec} by mathematical induction in the number $n$ of matrix multiplications.\\
{\em Induction basis:} For $n=1$ we have
\begin{subequations}
\begin{align}
\lvert A_{ii}\rvert&\leq\sum_{\substack{k\in\Lambda\\k\neq i}}\dfrac{\lvert\J\rvert}{\vertl i-k\vertr^\alpha}=|J|=U_1^{ii},\\
\lvert A_{ij}\rvert&\leq \dfrac{\lvert\J\rvert}{\vertl i-j\vertr^\alpha}= \dfrac{U_1^{ij}}{\vertl i-j\vertr^\alpha}\qquad \text{for $i\neq j$}.
\end{align} 
\end{subequations}
{\em Induction hypothesis:} Assume \eqref{e:rec} holds for some $n$.\\
{\em Inductive step:}
For $n+1$ and $i\neq j$, the left-hand side of \eqref{e:rec} can be bounded by
\begin{align}
\lefteqn{\Biggl\lvert\sum_{k_1,\hdots,k_{n}\in\Lambda}A_{jk_1}(t_1)\cdots A_{k_{n-1}k_{n}}(t_{n})A_{k_{n}i}(t_{n+1})\Biggr\rvert}\nonumber\\
&\underset{\eqref{e:rec}}{\leq}\sum_{\substack{k_{n}\in\Lambda\\k_{n}\neq j}}\dfrac{U_{n}^{jk_n}}{\vertl j-k_n\vertr^\alpha}\lvert A_{k_{n}i}(t_{n+1})\rvert+U_{n}^{jj}\lvert A_{ji}(t_{n+1})\rvert\nonumber\displaybreak[2]\\
&\leq U_{n}^\text{max}\sum_{\substack{k_{n}\in\Lambda\\k_{n}\neq i,j}}\dfrac{1}{\vertl j-k_n\vertr^\alpha}\dfrac{|\J|}{\vertl k_n-i\vertr^\alpha}\nonumber\\
&\quad+\dfrac{U_{n}^{ji}}{\vertl j-i\vertr^\alpha}\sum_{\substack{k_{n}\in\Lambda\\k_{n}\neq i}}\dfrac{ |\J|}{\vertl i-k_n\vertr^{\alpha}}+U_{n}^{jj}\dfrac{|\J|}{|i-j|^\alpha}\nonumber\\
&\underset{\eqref{e:repro}}{\leq}\dfrac{C_{ij}U_{n}^\text{max}+|J| U_{n}^{ji}+|\J| U_{n}^{jj}}{\vertl i-j\vertr^\alpha}=\dfrac{U_{n+1}^{ji}}{\vertl i-j\vertr^\alpha}.\label{e:Aij}
\end{align}
For $n+1$ and $i=j$, the matrix product is bounded by
\begin{eqnarray}
\lefteqn{\Biggl\lvert\sum_{k_1,\hdots,k_{n}\in\Lambda}A_{ik_1}(t_1)\cdots A_{k_{n-1}k_{n}}(t_{n})A_{k_{n}i}(t_{n+1})\Biggr\rvert}\nonumber\\
&\underset{\eqref{e:rec}}{\leq}&U_{n}^\text{max}\sum_{\substack{k_{n}\in\Lambda\\k_{n}\neq i}}\dfrac{1}{\vert i-k_n\vert^\alpha}\lvert A_{k_{n}i}(t_{n+1})\rvert+U_{n}^{ii}\lvert A_{ii}(t_{n+1})\rvert\nonumber
\\& \leq & C_{ii}U_{n}^\text{max}+\lvert J\rvert U_{n}^{ii}=U_{n+1}^{ii}.
\end{eqnarray}
This completes the proof of \eqref{e:rec} for all $n\geq 1$. 

Making use of the time-independent bound \eqref{e:rec}, the time-dependence of the integrand in \eqref{e:series} becomes trivial. Hence the integration can be performed by elementary means, yielding
\begin{equation}
\vertl\psii_j(t)\vertr\leq\dfrac{1}{\vertl i-j\vertr^\alpha}\sum_{n=1}^\infty \dfrac{U_n^{ij}t^{2n}}{(2n)!},
\end{equation}
which proves the stronger (middle) bound in (5c) for all $i,j\in\Lambda$ with $i\neq j$. (The case $i=j$ is not relevant here.) Here we have assumed convergence of the series in \eqref{e:series}, and the following calculation of the weaker bound in (5c) will confirm that this assumption is indeed justified.

An upper bound on $U_{n}^\text{max}$ is obtained by taking the supremum over pairs $i,j$ of the recursion relation (6),
\begin{align}
U_{n+1}^\text{max}&\leq \sup_{i,j\in\Lambda}\left(\lvert J\rvert U_{n}^{ij}+\lvert\J\rvert U_{n}^{ii}+C_{ij}U_{n}^\text{max}\right)\nonumber\\
&\leq U_{n}^\text{max}\biggl(\sup_{i,j\in\Lambda}C_{ij}+\lvert J+\J\rvert\biggr) = U_{n}^\text{max}v^2
\end{align}
which is solved by $U_{n}^\text{max}\leq v^{2n}$. Inserting the latter inequality into \eqref{e:Aij} yields
\begin{multline}
\Biggl\lvert\sum_{k_1,\hdots ,k_{n-1},k_{n}\in\Lambda}A_{jk_1}(t_1)\cdots A_{k_{n-1}k_{n}}(t_{n})A_{k_{n}i}(t_{n+1})\Biggr\rvert\\
\leq v^{2n+2}.
\end{multline}
Inserting this time-independent bound into \eqref{e:series} and performing the integration we obtain
\begin{equation}\label{e:ch1}
\vertl\psii_j(t)\vertr\leq\dfrac{1}{\vertl i-j\vertr^\alpha}\left(\sum_{n=1}^M \dfrac{(v t)^{2n}}{(2n)!}+\Vert \psi^{i}\Vert_\infty\dfrac{(v t)^{2(M+1)}}{(2(M+1))!}\right)
\end{equation}
for all $i,j\in\Lambda$ with $i\neq j$. The second term in the round brackets vanishes in the limit $M\to\infty$. This implies convergence of the series, and we obtain
\begin{equation}
\vertl\psii_j(t)\vertr\leq\frac{\cosh\left(v t\right)-1}{\vertl i-j\vertr^\alpha},
\end{equation}
which proves the weaker (rightmost) bound in (5c).

\section{B. Bounds for general classical long-range systems}
\setcounter{section}{2}
\setcounter{equation}{0}
\setcounter{figure}{0}


As a lattice, we consider a set of vertices $\Lambda$ and a set of edges $E$ connecting pairs of vertices. The graph distance $d(i,j)$ (number of edges of the shortest path connecting the sites $i,j\in\Lambda$) serves as a metric on the graph $(\Lambda,E)$.

To each vertex $i\in\Lambda$ we assign a $\mu$-dimensional manifold $M_i$ as the local configuration space. 
For any finite subset $X\subset\Lambda$, the configuration space associated with $X$ is the product space $\YY_X=\prod_{i\in X}M_i$, and the phase space associated with $X$ is the cotangent bundle $\XX_X = T^*(\YY_X)$.
We define the Hamiltonian $H_\Lambda:\XX_\Lambda\to\RR$ as a real function on phase space, and, for given initial conditions, it generates a flow on $\XX_\Lambda$ in the usual way via Hamilton's equations.
Within this setting, following Section 2 of \cite{Marchioro_etal78}, one can show that Eqs.\ (1) and (2) of the main text hold for differentiable functions $f,g:\XX_\Lambda\to\RR$ with bounded derivatives.

We consider Hamiltonian functions of a standard form,
\begin{equation}\label{e:H_gen}
H_\Lambda=\sum_{i\in\Lambda}\frac{p_i^2}{2m_i}+\mathscr{N}_\Lambda \sum_{X\subset\Lambda}\Phi^X_\Lambda\left(\{q_k\}_{k\in X}\right),
\end{equation}
consisting of a kinetic term quadratic in the momenta $p_i\in T^*(M_i)$, and a general interaction term depending on the coordinates $q_i\in M_i$ via the differentiable $n$-vertex interactions $\Phi_\Lambda^X:\YY_X\to\RR$, with $n=|X|$ the number of elements in the set $X$. The normalization factor $\mathscr{N}_\Lambda$ is chosen such that $H_\Lambda$ is extensive \footnote{We allow both, $\Phi_\Lambda^X$ and $\mathscr{N}_\Lambda$, to explicitly depend on $\Lambda$. This can be necessary when, for example, 2-body and 3-body interactions are present on a $D$-dimensional regular lattice and both interactions decay like $r^{-\alpha}$ with the distance $r$ and some exponent $\alpha$ smaller than $D$. In this case the 2- and 3-body terms require different normalization factors in order to guarantee extensivity of the Hamiltonian, and this difference can be accounted for by means of a $\Lambda$-dependent 3-body interaction $\Phi_\Lambda^{\{i,j,k\}}$.}.

For proving a Lieb-Robinson-type bound, we require the interactions $\Phi_\Lambda^X$ to satisfy
\begin{subequations}
\begin{align}
\sup_{i,j\in\Lambda}\sum_{X\ni i,j}\sup_{q\in\YY_X}\frac{\left|\dfrac{\partial^2 \Phi_\Lambda^X}{\partial\q_i\partial\q_j}\left(\{q_k\}_{k\in X}\right)\right|}{ F(d(i,j))}&<\infty\quad\text{for $i\neq j$},\label{eG:condition1}\\
\mathscr{N}_\Lambda\sup_{i\in\Lambda}\sum_{X\ni i}\sup_{q\in\YY_X}\frac{\left|\dfrac{\partial^2 \Phi_\Lambda^X}{\partial\q_i^2}\left(\{q_k\}_{k\in X}\right)\right|}{ F(d(i,i))}&<\infty.\label{eG:condition2}
\end{align}
\end{subequations}
For local configuration space manifolds $M_i$ of dimensions $\mu$ greater than 1, $\partial^2 \Phi_\Lambda^X/(\partial\q_i\partial\q_j)$ is a $\mu\times\mu$-matrix. Here and for the remainder of this section, the vertical bars $\lvert\cdot\rvert$ denote an elementwise matrix maximum norm, $\vert B\rvert=\max_{i,j}\lvert B_{ij}\rvert$ for a matrix $B$ with elements $B_{ij}$. Similar to Section 1.1 of \cite{NachtergaeleOgataSims06}, the non-increasing function $F:[0,\infty)\to(0,\infty)$ has the following properties:
\begin{enumerate}
\renewcommand{\labelenumi}{(\roman{enumi})}
\item $\mathscr{N}_\Lambda F$ is uniformly summable over $\Lambda$, i.e.,
\begin{equation}
\lVert F_\Lambda\rVert:=\sup_{i\in\Lambda}\sum_{j\in\Lambda}\mathscr{N}_\Lambda F(d(i,j))<\infty,
\end{equation}
\item $F$ satisfies
\begin{equation}
C_\Lambda=\mathscr{N}_\Lambda\sup_{i,j\in\Lambda}\sum_{\substack{k\in\Lambda\\k\neq i}}\frac{F(d(i,k))F(d(k,j))}{F(d(i,j))}<\infty.
\label{eG:repro}
\end{equation}
\end{enumerate}

Within this setting, we obtain the following bounds on the partial derivatives on the right-hand side
of (2),
\begin{subequations}
\begin{align}
\left\lvert\frac{\partial \q_j(t)}{\partial \q_i(0)}\right\rvert 
&\leq \frac{F(d(i,j))}{\mu m_j}\sum_{n=1}^\infty \dfrac{\mu^n V_n^{ij}t^{2n}}{(2n)!}\nonumber\\
&\leq \frac{F(d(i,j))}{\mu m_j}\left (\cosh\left (v t\right )-1\right ),\label{e:pq1_gen}\\
\left\lvert\frac{\partial \q_j(t)}{\partial \p_i(0)}\right\rvert
&\leq \frac{F(d(i,j))}{\mu m_j}\sum_{n=1}^\infty \dfrac{\mu^n V_n^{ij}t^{2n+1}}{(2n+1)!}\nonumber\\
&\leq \frac{F(d(i,j))}{\mu v m_j}\left (\sinh\left (v t\right )-v t\right ),\label{e:pq2_gen}\\
\left\lvert\frac{\partial \p_j(t)}{\partial \q_i(0)}\right\rvert
&\leq \frac{F(d(i,j))}{\mu m_j}\sum_{n=1}^\infty \dfrac{\mu^n V_n^{ij}t^{2n}}{(2n)!}\nonumber\\
&\leq v \frac{F(d(i,j))}{\mu m_j}\left (\sinh\left (v t\right )\right ),\label{e:pq3_gen}\\
\left\lvert\frac{\partial \p_j(t)}{\partial \p_i(0)}\right\rvert 
&\leq \frac{F(d(i,j))}{\mu m_j}\sum_{n=1}^\infty \dfrac{\mu^n V_n^{ij}t^{2n}}{(2n)!}\nonumber\\
&\leq \frac{F(d(i,j))}{\mu m_j}\left (\cosh\left (v t\right )-1\right ), \label{e:pq4_gen}
\end{align}
\end{subequations}
with
\begin{equation}
v=\sqrt{\mu\left[ C_\Lambda+F(0)\right ]}.
\end{equation}
The coefficients $V_n^{ij}$ are defined recursively by
\begin{equation}\label{e:Vn}
V_{n+1}^{ij}=C_{ij}V_n^{\text{max}}+F(0)V_n^{ij}
\end{equation}
with $V_1^{ii}=1$ and $V_1^{ij}=\mathscr{N}_\Lambda$ for $i\neq j$,
\begin{align}
V_n^{\text{max}}&=\sup_{i,j\in\Lambda} V_n^{ij},\\
C_{ij}&=\mathscr{N}_\Lambda\sum_{\substack{k\in\Lambda\\k\neq i}}\frac{F(d(i,k))F(d(k,j))}{F(d(i,j))}.
\label{eG:Cij}
\end{align}


{\em Proof.---}The equations of motion corresponding to the Hamiltonian \eqref{e:H_gen} are
\begin{subequations}
\begin{align}
\dfrac{\text{d}}{\text{d} t}\p_j(t)&=-\mathscr{N}_\Lambda\sum_{X\ni j} \derp{\Phi^X_\Lambda}{\q_j}\left(\{\q_k(t)\}_{k\in X}\right),
\label{eG:Homega}\\
\dfrac{\mathrm{d}}{\text{d} t}\q_{j}(t)&=\frac{\p_{j}(t)}{m_j}.
\label{eG:Hq_}
\end{align}
\end{subequations}
Integrating over $t$ and take the partial derivative with respect to $\q_i(0)$ or $p_i(0)$, we obtain
\begin{subequations}
\begin{align}
\derp{\q_{j}(t)}{\q_i(0)}=&\delta_{ij}\id_\mu+\frac{1}{m_j}\intt\derp{\p_{j}(s)}{\q_{i}(0)}\ds,\label{eG:tt}\\
\derp{\q_{j}(t)}{\p_{i}(0)}=&\frac{1}{m_j}\intt\derp{\p_{j}(s)}{\p_{i}(0)}\ds,\label{eG:to}\\
\derp{\p_{j}(t)}{\q_{i}(0)}=&-\mathscr{N}_\Lambda\intt \sum_{X\ni j} \derp{}{\q_i(0)}\derp{\Phi^X_\Lambda(\{\q_k(s)\}_{k\in X})}{\q_j}\ds \nonumber\\
=&-\mathscr{N}_\Lambda\intt\sum_{X\ni j}\sum_{k\in X} \frac{\partial^2 \Phi^X_\Lambda}{\partial \q_j\partial\q_k}\frac{\partial \q_k(s)}{\partial \q_i(0)}\ds \nonumber\\
=&\intt\sum_{k\in\Lambda}\mathscr{A}_{jk}\frac{\partial \q_k(s)}{\partial \q_i(0)}\ds ,\label{eG:ot} \\
\derp{\p_{j}(t)}{\p_i(0)}=&\delta_{ij}\id_\mu+\intt\sum_{k\in\Lambda}\mathscr{A}_{jk}\frac{\partial \q_k(s)}{\partial \p_i(0)}\ds, \label{eG:oo}
\end{align}
\end{subequations}
where $\id_\mu$ denotes the $\mu\times\mu$ identity matrix and
\begin{equation}\label{eG:scrA}
\mathscr{A}_{jk}=-\mathscr{N}_\Lambda\sum_{X\ni j,k} \frac{\partial^2 \Phi^X_\Lambda}{\partial \q_j\partial\q_k}.
\end{equation}
Inserting \eqref{eG:ot} into \eqref{eG:tt} and introducing the definition
\begin{equation}
\psii=\left(\psii_{j}\right)_{j\in\Lambda}\qquad\text{with}\quad\psii_{j}(t)=\derp{q_j(t)}{q_i(0)}
\end{equation}
we obtain
\begin{equation}\label{eG:dintegral}
\psii_{j}(t)=\delta_{ij}\id_\mu+\intt\int_0^{t_1} \left[\mathscr{A}\psii\right]_j\!(t_2)\,\dt_2\,\dt_1.
\end{equation}
This equation is formally identical to \eqref{e:dintegral}, the only difference being that each entry of the $|\Lambda|\times|\Lambda|$-matrix $\mathscr{A}$ and of the $|\Lambda|$-vector $\psii$ is now a $\mu\times\mu$-matrix. Integrating \eqref{eG:dintegral} by parts and $M$-fold iteration of the resulting expression yields
\begin{multline}
\psii_j(t)=\delta_{ij}\id_\mu+\sum_{m=1}^M \biggl(\intt\int^{t_1}_0\cdots\int^{t_{m-1}}_0\dt_m\cdots\dt_1\\
\times(t-t_1)\cdots(t_{m-1}-t_m)\left[\mathscr{A}(t_1)\cdots \mathscr{A}(t_m)\deltai\right]_{j}\biggr)\\
+\intt\int^{t_1}_0 \cdots\int^{t_{M}}_0 
(t-t_1)\cdots(t_{M}-t_{M+1})\\
\times\left[\mathscr{A}(t_1)\cdots \mathscr{A}(t_{M+1})\psii(t_{M+1})\right]_{j}\dt_{M+1}\cdots\dt_1.
\label{eG:series}
\end{multline}

Analogous to Appendix A we prove by induction the inequality
\begin{equation}\label{eG:recA}
  \Biggl\lvert\sum_{k_1,\dotsc,k_{n-1}\in\Lambda}\mathscr{A}_{jk_1}(t_1)\cdots \mathscr{A}_{k_{n-1}i}(t_n)\Biggr\rvert\leq \mu^{n-1}F(d(i,j))V_{n}^{ij}
\end{equation}
for $i,j\in\Lambda$, with coefficients $V_{n}^{ij}$ defined in \eqref{e:Vn}.\\
{\em Induction basis:} Making use of the conditions \eqref{eG:condition1} and \eqref{eG:condition2}, we find
\begin{equation}
\lvert \mathscr{A}_{ij}\rvert\leq F(d(i,j))\begin{cases}\mathscr{N}_\Lambda & \text{for $i\neq j$},\\1 & \text{for $i=j$}.\end{cases}
\end{equation}
Hence for $n=1$ we have
\begin{equation}
\lvert \mathscr{A}_{ij}\rvert\leq F(d(i,j))V_1^{ij} \qquad\forall i,j\in\Lambda.
\end{equation}
{\em Induction hypothesis:} Assume \eqref{eG:recA} holds for some $n$.\\
{\em Inductive step:}
For $n+1$, the left-hand side of \eqref{eG:recA} can be bounded by
\begin{eqnarray}
\lefteqn{\Biggl\lvert\sum_{k_1,\hdots,k_{n}\in\Lambda}\mathscr{A}_{jk_1}(t_1)\cdots \mathscr{A}_{k_{n-1}k_{n}}(t_n)\mathscr{A}_{k_{n}i}(t_{n+1})\Biggr\rvert}\nonumber\\
&\underset{\eqref{eG:recA}}{\leq}&\mu^{n-1}\sum_{\substack{k_{n}\in\Lambda}}V_{n}^{k_n j}F(d(k_n,j))\lvert \mathscr{A}_{k_n i}(t_{n+1})\rvert\nonumber\\
&\leq & \mu^{n} F(d(i,i))F(d(i,j))V_n^{ij}\nonumber\\
&&+ \mu^{n} \mathscr{N}_\Lambda\sum_{\substack{k_n\in\Lambda\\k_n\neq i}}F(d(i,k_n))F(d(k_n,j))V_n^{k_n j}\nonumber\\
&\underset{\eqref{eG:Cij}}{\leq}&\mu^{n} F(d(i,j))V_{n+1}^{ij},\label{eG:Aij}
\end{eqnarray}
where the powers of $\mu$ arise from the the fact that the elements $\mathscr{A}_{ij}$ are $\mu\times\mu$-matrices. This completes the proof of \eqref{eG:recA} for all $n\geq 1$. 

Inserting this bound into \eqref{eG:series} and performing the resulting trivial integral we obtain \eqref{e:pq1_gen}. The bounds in \eqref{e:pq2_gen}--\eqref{e:pq4_gen} are obtained along the same lines.

\section{C. Large-system limit of the constants \texorpdfstring{$C_{ij}$}{Cij}}
\setcounter{section}{3}
\setcounter{equation}{0}
\setcounter{figure}{0}

\subsection{C.1 Discussion of condition \texorpdfstring{\eqref{e:repro}}{(A.8)}}
For the proof of \eqref{e:rec} we require that
\begin{equation}\label{e:repro2}
C_{ij}<\infty\qquad\forall i,j\in\Lambda,~i\neq j
\end{equation}
with $C_{ij}$ as defined in (8). This condition is similar to Eq.\ (2.3) of Ref.\ \cite{HastingsKoma06}, used for the proof of a Lieb-Robinson-type bound for long-range interacting quantum systems, but our condition differs by the additional factor $\J$ in (8). For $\alpha>d$, $\J$ converges to a nonzero constant in the thermodynamic limit, and our condition \eqref{e:repro2} becomes identical to Eq.\ (2.3) of Ref.\ \cite{HastingsKoma06}. For $\alpha\leq d$, however, $\J$ vanishes in the thermodynamic limit, canceling the divergence that---as we will show next---would otherwise occur in \eqref{e:repro2}.

\begin{figure}\centering
\includegraphics[width=0.96\linewidth]{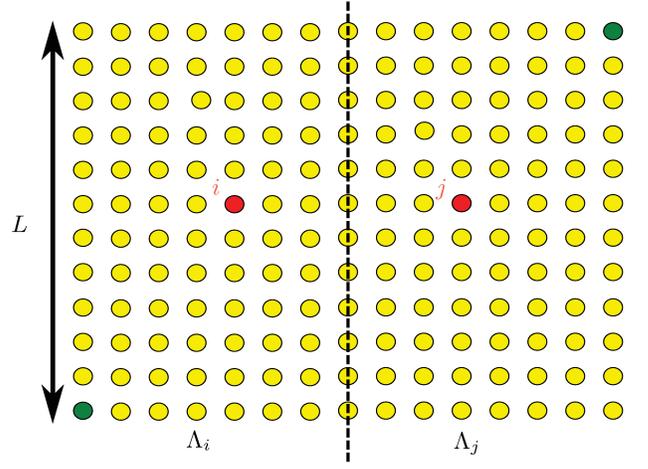}
\caption{\label{f:latticedem} Sketch of a two-dimensional square lattice $\Lambda$, divided into two regions $\Lambda_i$ and $\Lambda_j$ at distance $|i-j|/2$ of the spins $i$ and $j$ in red. The green (darker) corner sites denote two spins separated by the largest possible distance $|i-j|=2L$ on the lattice.}
\end{figure}

For simplicity, we sketch the existence proof of a finite upper bound on $C_{ij}$ for the case $i\neq j$ on a two-dimensional square lattice, but the generalization to other lattices and higher dimensionality is straightforward. As illustrated in Fig.\ \ref{f:latticedem}, we divide $\Lambda$ into two parts $\Lambda_i$ and $\Lambda_j$, such that the (dashed) line separating the two parts is perpendicular to the line connecting $i$ and $j$, and centered between the two sites. For simplicity we assume a reflection symmetric arrangement and write
\begin{equation}
C_{ij}\leq 2|\J| \sum_{k\in\Lambda_i\setminus\{i\}}\frac{|i-j|^\alpha}{|i-k|^\alpha|j-k|^\alpha},
\end{equation}
but generalizations are again straightforward. By construction, every site $k\in\Lambda_i$ is at least a distance $|i-j|/2$ away from $j$, and this implies the bound
\begin{equation}
C_{ij}\leq  2^{\alpha+1}|\J| \sum_{k\in\Lambda_i\setminus\{i\}}\frac{1}{|i-k|^\alpha}\leq2^{\alpha+1}|J|,
\end{equation}
which remains finite in the thermodynamic limit. Similarly, $C_{ii}$ can be shown to go to zero in that limit.

While such a finite upper bound on $C_{ij}$ is necessary for the proof in Appendix A, a nonvanishing lower bound is also of interest as it ensures that the $U_n^{ij}$-modification of the spatial dependence $|i-j|^{-\alpha}$ does not fade away in the thermodynamic limit. For $\alpha>d$, both $J_\Lambda$ and $C_{ij}$ converge to a finite, nonzero value (see also Appendix B.2), and the existence of a nonzero lower bound on $C_{ij}$ follows immediately. To prove a bound for $\alpha\leq d$, we note that $dL$ is the maximum distance that can occur on a square lattice patch of linear dimension $L$ (see Fig.\ \ref{f:latticedem}), implying $\protect{\vert j-k\rvert}\leq dL$ for all $k\in\Lambda$. Inserting this inequality into (8), we obtain the lower bound
\begin{equation}
C_{ij}\geq\left(\frac{|i-j|}{dL}\right)^{\!\alpha} |\J| \!\sum_{k\in\Lambda\setminus\{i,j\}}\!\!|i-k|^{-\alpha}=c_i |J|\left(\frac{|i-j|}{dL}\right)^{\!\alpha}
\end{equation}
with some nonzero, $i$-dependent proportionality constant $c_i$ that approaches unity in the thermodynamic limit. From this expression, one can read off that $C_{ij}$ goes to zero in the large-system limit for any fixed pair of lattice sites $i,j\in\Lambda$. A nonzero lower bound is obtained only when considering a sequence of $C$-coefficients with a fixed ratio $r=|i-j|/L$,
\begin{equation}
C_r\geq c_i |J|\left(\frac{r}{d}\right)^\alpha.
\end{equation}
Hence, for $\alpha\leq d$ and a fixed $r>0$ we have
\begin{equation}
0<|J|\left(\frac{r}{d}\right)^\alpha\leq C_r\leq |J|2^{\alpha+1}<\infty
\end{equation}
in thermodynamic limit.

\subsection{C.2 Large-system asymptotics}

Both, the weaker and the sharper bounds in (5a)--(5d) may inherit a characteristic system-size dependence through the dependence of the constants $\J$ and $C_{ij}$ on the number $N=|\Lambda|$ of lattice sites. Here we analyze this dependence for the (one-dimensional) $\alpha XY$ chain in the asymptotic regime of large $N$ for three different regimes of the long-range exponent $\alpha$. This analysis is similar to Appendix B.1, but focuses on asymptotic behavior instead of upper and lower bounds. 

\paragraph{\texorpdfstring{$\alpha>1$}{alpha>1}:}
Due to the translation invariance of the $\alpha XY$ chain with periodic boundary conditions, we can write the lattice-dependent coupling constant (4) in the large-system limit as
\begin{equation}\label{e:JZZ}
J_\ZZ=J \Big/\sum_{j=1}^\infty \frac{2}{j^\alpha} = \frac{J}{2\zeta(\alpha)},
\end{equation}
where $\zeta$ denotes the Riemann zeta function and $\alpha>1$ is required for convergence of the sum. The sum
\begin{equation}\label{e:sum}
\sum_{k\in\mathbb{Z}\setminus\{i,j\}}\frac{|i-j|^\alpha}{|i-k|^\alpha|j-k|^\alpha}
\end{equation}
in the definition (8) of the constants $C_{ij}$ also converges for all $\alpha>0$ in the infinite-size limit \footnote{See also Section III of the Supplemental Material accompanying \cite{HaukeTagliacozzo13}} and can be evaluated numerically to high precision. By interpreting it as a Riemann sum it is also possible to bound the error of such a numerical evaluation in terms of hypergeometric functions.

\paragraph{\texorpdfstring{$1/2<\alpha<1$}{1/2<alpha<1}:}
For $\alpha<1$, the sum in \eqref{e:JZZ} does not converge in the limit of infinite chain length $N\to\infty$. We can determine the asymptotic large-$N$ behavior by writing
\begin{equation}
J_\ZZ=J \Big/\sum_{j=1}^{N/2} \frac{2}{j^\alpha} = \frac{JN^{\alpha-1}}{2}\Big/\Biggl(\frac{1}{N}\sum_{j=1}^{N/2}\frac{1}{|j/N|^{\alpha}}\Biggr).
\end{equation}
Interpreting the denominator of this expression as a Riemann sum, we obtain
\begin{equation}
J_\ZZ\sim \frac{JN^{\alpha-1}}{2}\Big/ \int_0^{1/2}\dd x\,x^{-\alpha}= \frac{(1-\alpha)JN^{\alpha-1}}{2^\alpha},
\end{equation}
valid asymptotically for large $N$. In the limit $N\to\infty$, $J_\ZZ$ goes to zero.

The asymptotic behavior of $C_{ij}$ depends on the precise way in which the limit is taken. If the large-$N$ limit is considered for $\alpha>1/2$ and some fixed lattice sites $i$ and $j$ [as in the context of the recursion relation (6)], the sum \eqref{e:sum} converges. Since $\J$ in the definition (8) goes to zero, $C_{ij}$ also vanishes. Considering however, instead of fixed sites $i$ and $j$, a fixed ratio $r=|i-j|/N$, $C_{ij}$ converges in the large-$N$ limit to a finite, nonzero value. This becomes evident from the integral representation
\begin{equation}\label{e:C_tdl}
C_{ij}\sim\frac{(1-\alpha)|J|}{2^\alpha}\left|\frac{i-j}{N}\right|^\alpha\int_{-1/2}^{1/2} \frac{dy}{|y|^\alpha}\left|\left|\frac{i-j}{N}\right|-y\right|^{-\alpha},
\end{equation}
where $|\cdot|$ denotes the shortest distance along a circle of unit circumference, i.e.,
\begin{equation}
|x|=\begin{cases} x+1 & \text{for $-1\leq x<-1/2$,}\\ -x &  \text{for $-1/2\leq x<0$,}\\ x &  \text{for $0\leq x<1/2$,}\\ 1-x &  \text{for $1/2\leq x<1$.}\end{cases}
\end{equation}
This is the relevant asymptotic behavior when considering the supremum over all $C_{ij}$, as in the definition (7) of $v$ or (the upper bound on) $U_n^\text{max}$. 

\paragraph{\texorpdfstring{$0<\alpha<1/2$}{0<alpha<1/2}:}
By similar techniques, one finds that the sum \eqref{e:sum} diverges like $N^{1-2\alpha}$ for fixed values of $i$ and $j$, and $C_{ij}$ therefore vanishes like $N^{-\alpha}$. The scaling of $J_\ZZ$, $v$, and $U_n^\text{max}$ remains unchanged from the case $1/2<\alpha<1$.

This discussion establishes the three different scaling re\-gimes relevant for the coefficients $J_\ZZ$, $C_{ij}$, $v$, and $U_n^\text{max}$ occurring in the bounds (5a)--(5d). The threshold values $\alpha=1/2$ and $\alpha=1$ at which the switchings from one regime to another occur coincide with the threshold values observed for the equilibration times in Ref.\ \cite{BachelardKastner13}.

\section{D. Details of the numerical simulations}
\setcounter{section}{4}
\setcounter{equation}{0}
\setcounter{figure}{0}

\begin{figure}\centering
\includegraphics[width=0.96\linewidth]{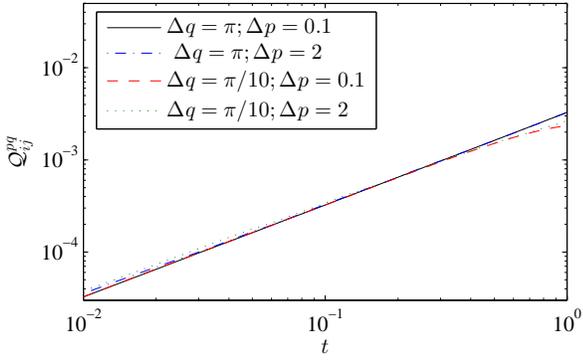}
\caption{\label{f:initialconditions}%
Temporal behavior of perturbations in the $\alpha XY$ chain for different initial conditions. The numerical data are for chains of length $N=4096$ and long-range exponent $\alpha=1/2$. The curves show $\mathcal{Q}_{ij}^{pq}$ as a function of time for initial conditions with different values of $\Delta q$ and $\Delta p$.
}%
\end{figure}

For the numerical simulations reported in this paper we used initial conditions with particle positions randomly drawn from a flat distribution over the interval $[-\Delta q,\Delta q]$, and momenta from $[-\Delta p,\Delta p]$. While the bounds (5a)--(5d) are independent of the initial conditions, it is not clear that the same is true for the actual spreading of perturbations. To investigate this issue, we analyze to what extend different values of $\Delta q$ and $\Delta p$ affect the spreading of perturbations. The plot in Fig.\ \ref{f:initialconditions} illustrates that the effect of different initial conditions is very weak, as the amplitude of the growth of $\mathcal{Q}_{ij}^{pq}$ is almost unaffected, and the results for other difference quotients are similar. Gaussian initial conditions were also tested, and gave analogous growth rates. This corroborates that the various features of the spreading of perturbations discussed in the main text are essentially independent of the choice of initial conditions.

While for all initial conditions studied the numerical results comply with the bounds (5a)--(5d), the fitting of the function (11) works better, and for longer times, for initial states with a small value of $\Delta$. We have used initial conditions drawn from a distribution with $\Delta q=\pi$ and $\Delta p=0.1$ throughout the paper, unless otherwise stated.

\bibliography{LRLR}

\end{document}